\documentclass[%
 reprint,
superscriptaddress,
 amsmath,amssymb,
 aps,prl,
floatfix,
]{revtex4-2}

\usepackage{graphicx}
\usepackage{dcolumn}
\usepackage{bm}
\usepackage{braket}
\usepackage{hyperref}
\hypersetup{pdfpagemode=FullScreen,
colorlinks=true, citecolor = blue, linkcolor=blue, filecolor=blue}
\usepackage[mathlines]{lineno}
\usepackage{amsmath}
\usepackage[T1]{fontenc}
\usepackage{color}

\usepackage[normalem]{ulem}

\begin{document}

\preprint{APS/123-QED}

\title{Detecting Axion Dark Matter with an Organic Molecular Maser}

\author{Hongliang Wu}%
\altaffiliation[]{These authors contributed equally.}
\affiliation{%
Key Laboratory of Advanced Optoelectronic Quantum Architecture and Measurements of Ministry of Education, Center for Interdisciplinary Science of Optical Quantum and NEMS Integration, School of Physics, Beijing Institute of Technology, Beijing 100081, China
}%
\affiliation{%
School of Integrated Circuits and Electronics, Beijing Institute of Technology, Beijing 100081, China
}%

\author{Yuchen Han}%
\altaffiliation[]{These authors contributed equally.}
\affiliation{%
Key Laboratory of Advanced Optoelectronic Quantum Architecture and Measurements of Ministry of Education, Center for Interdisciplinary Science of Optical Quantum and NEMS Integration, School of Physics, Beijing Institute of Technology, Beijing 100081, China
}%

\author{Zhengtao Wang}%
\affiliation{%
Key Laboratory of Advanced Optoelectronic Quantum Architecture and Measurements of Ministry of Education, Center for Interdisciplinary Science of Optical Quantum and NEMS Integration, School of Physics, Beijing Institute of Technology, Beijing 100081, China
}%

\author{Dezhi Zheng}
\affiliation{%
State Key Laboratory of Environment Characteristics and Effects for Near-space, Beijing Institute of Technology, Beijing 100081, China
}%
\affiliation{%
MIIT Key Laboratory of Complex-field Intelligent Sensing, Beijing Institute of Technology, Beijing 100081, China
}%

\author{Yeliang Wang}
\affiliation{%
School of Integrated Circuits and Electronics, Beijing Institute of Technology, Beijing 100081, China
}%
\affiliation{%
MIIT Key Laboratory for Low-Dimensional Quantum Structure and Devices, Beijing Institute of Technology, Beijing 100081, China
}%

\author{Liu Yang}
\affiliation{%
Key Laboratory of Advanced Optoelectronic Quantum Architecture and Measurements of Ministry of Education, Center for Interdisciplinary Science of Optical Quantum and NEMS Integration, School of Physics, Beijing Institute of Technology, Beijing 100081,  China
}%

\author{Zhiwei Wang}
\affiliation{%
Key Laboratory of Advanced Optoelectronic Quantum Architecture and Measurements of Ministry of Education, Center for Interdisciplinary Science of Optical Quantum and NEMS Integration, School of Physics, Beijing Institute of Technology, Beijing 100081, China
}%
\affiliation{%
Beijing Key Lab of Nanophotonics and Ultrafine Optoelectronic Systems, Beijing Institute of Technology, Beijing 100081, China.
}%
\affiliation{%
International Centre for Quantum Materials, Beijing Institute of Technology, Zhuhai 519000, China.
}%

\author{Bo Zhang}
\email{bozhang_quantum@bit.edu.cn}
\affiliation{%
Key Laboratory of Advanced Optoelectronic Quantum Architecture and Measurements of Ministry of Education, Center for Interdisciplinary Science of Optical Quantum and NEMS Integration, School of Physics, Beijing Institute of Technology, Beijing 100081, China
}%
\affiliation{%
State Key Laboratory of Environment Characteristics and Effects for Near-space, Beijing Institute of Technology, Beijing 100081, China
}%
\affiliation{%
MIIT Key Laboratory of Complex-field Intelligent Sensing, Beijing Institute of Technology, Beijing 100081, China
}%
\affiliation{%
Center for Quantum Technology Research and Key Laboratory of Advanced Optoelectronic Quantum Architecture and Measurements (MOE), Beijing Institute of Technology, Beijing 100081, China
}%

\author{Dmitry Budker}
\email{budker@uni-mainz.de}
\affiliation{%
Johannes Gutenberg-Universität Mainz, 55128 Mainz, Germany
}%
\affiliation{%
Helmholtz-Institut, GSI Helmholtzzentrum für Schwerionenforschung, 55128 Mainz, Germany
}%
\affiliation{%
Department of Physics, University of California, Berkeley, California 94720, USA
}%

\author{Jun Zhang}
\email{zhjun@bit.edu.cn}
\affiliation{%
State Key Laboratory of Environment Characteristics and Effects for Near-space, Beijing Institute of Technology, Beijing 100081, China
}%
\affiliation{%
MIIT Key Laboratory of Complex-field Intelligent Sensing, Beijing Institute of Technology, Beijing 100081, China
}%
\affiliation{%
State Key Laboratory of CNS/ATM, Beijing Institute of Technology, Beijing 100081, China
}%

\begin{abstract}
We present a novel quantum sensing approach to search for axion–electron interactions around the axion mass of 6\,$\mu$eV. In this region, laboratory searches are relatively scarce, and our direct experiment measuring the axion–electron coupling constant reaches the sensitivity of $8 \times 10^{-6}\,\rm{GeV}^{-1}$. The method, based on an organic molecular maser establishes a proof-of-principle for quantum-enhanced detection, with a corresponding magnetic field sensitivity of $0.85\,\rm{fT/\sqrt{Hz}}$. The methodology is generic and can be readily extended to other physical systems, further broadening its applicability in quantum sensing and dark matter searches.
\end{abstract}

\maketitle

\section{introduction}
The existence of dark matter was first evidenced decades ago through observations of the high velocities of galaxies within massive clusters \cite{RevModPhys.90.045002}. Subsequent experimental results have shown that dark matter constitutes approximately 26\% of the total energy density of the Universe. Unraveling its composition would significantly advance the fields of astrophysics and cosmology \cite{doi:10.1126/science.aal3003,RevModPhys.90.025008}. Weakly Interacting Massive Particles (WIMPs) were once among the most theoretically favored dark matter candidates. However, a growing number of experiments have increasingly constrained and are now largely excluding this possibility \cite{PhysRevLett.112.091303}. The Quantum Chromodynamics (QCD) axion \cite{PhysRevLett.38.1440}, originally proposed to resolve the strong CP problem, has emerged as a compelling candidate for dark matter \cite{PRESKILL1983127,RevModPhys.82.557}. Additionally, axion-like particles (ALPs), which share similar pseudoscalar properties \cite{annurev:/content/journals/10.1146/annurev.nucl.012809.104433}, have also attracted considerable attention in recent research.

Axions are theorized to interact with standard-model particles through several couplings, including axion–photon \cite{Anastassopoulos2017,PhysRevD.97.092001,PhysRevLett.104.041301,PhysRevLett.124.101303,PhysRevLett.120.151301,PhysRevD.96.123008}, axion–gluon \cite{PhysRevX.4.021030}, and axion–fermion \cite{PhysRevLett.113.161801,BogdanA.Dobrescu_2006,NI1994153} couplings. Considerable progress has been made in probing the axion–photon coupling, particularly in the $\mu$eV mass range where axions are plausible cold dark matter candidates \cite{TURNER199067}. A number of haloscope experiments—such as those carried out by IBS and collaborations between CNRS and IBS—are steadily approaching the theoretical sensitivity limit space \cite{PhysRevLett.124.101802,PhysRevLett.130.071002,10.3389/fphy.2024.1358810}. In contrast, axion–fermion couplings remain far less explored despite their ability to induce phenomena at microscopic scales such as magnetic moment modulation or spin transitions \cite{PhysRevD.88.035023,PhysRevX.4.021030}. Recent studies have attempted to probe axion-nucleon interactions with high-sensitivity nuclear spin-based quantum sensors \cite{Jiang2021,doi:10.1126/sciadv.abi9535,PhysRevLett.129.051801,PhysRevLett.133.191801,Jiang_2025}. However, laboratory searches directly targeting axion–electron couplings in the $\mu$eV mass regime are relatively scarce \cite{PhysRevLett.124.171801}, leaving this parameter space largely unexplored.

In this study, we introduce an organic molecular maser as a spin-based quantum sensor to probe the axion–electron couplings. By exploiting the coupling between cosmic axions and electron spins to induce stimulated emission, the scheme converts the weak axion signal into a measurable electromagnetic signal and subsequently amplifies it, providing a direct experimental pathway to measure axion–electron couplings for axion masses near 6\,$\mu$eV. In this parameter space, we report a direct laboratory constraint, reaching a sensitivity to the axion–electron coupling constant of $8 \times 10^{-6}\,\rm{GeV}^{-1}$ (95\% confidence level), assuming that the axions with this particular mass constitute all of the local galactic dark matter. The corresponding sensitivity to an oscillating magnetic field is $0.85\,\rm{fT/\sqrt{Hz}}$. Our measurement is sensitive to the amplitude of the projection of the induced pseudo-magnetic field on the sensitive axis of the maser. 

Notably, this method circumvents the need for ultra-strong magnets and cryogenic environments typically required in axion–photon conversion experiments, as well as the complex control pulse sequences demanded by many other quantum sensors \cite{doi:10.1126/sciadv.abq8158}. Moreover, the methodology is generic and can be extended to other spin systems.

\section{principle}

Axions and ALPs can interact with fermions, and the corresponding Lagrangian can be written as \cite{Bloch2020}:
\begin{align}
     L = g_{aff}\overline{\psi}_{f}\gamma^{\mu}\gamma^{5}{\psi}_{f}\partial_{\mu}a \label{eq.1},
\end{align}
where $a$ and $f$ stand for the fields of axion and fermion, respectively, and $g_{aff}$ represents the axion–fermion coupling constant. In this article, we employ electron spins to detect axions, and therefore we replace $f$ with $e$ in subsequent expressions. In the non-relativistic limit, this interaction resembles the magnetic interaction between spin and an external magnetic field in the Standard Model. Consequently, an axion-induced oscillating pseudo-magnetic field can be defined, whose Compton frequency $\nu_{a}$ is determined by the axion mass: $\nu_{a} = m_{a}c^{2}/h$, where $c$ is the speed of light, $h$ is Planck’s constant and $m_{a}$ is mass of the axion. Based on this, we can construct the Hamiltonian describing the interaction between the axion field and the electron spins \cite{Bloch2020,Garcon_2017,PhysRevD.89.043522}:
\begin{align}
     H = g_{aee}\sqrt{2\hbar^{3}c\rho_{\rm{DM}}}sin(2\pi\nu_{a}t)\textbf{v}\cdot \bm{\sigma}_{e} \label{eq.2}.
\end{align}

\begin{figure}[htbp!]
\centering
\includegraphics[width=8.8cm]{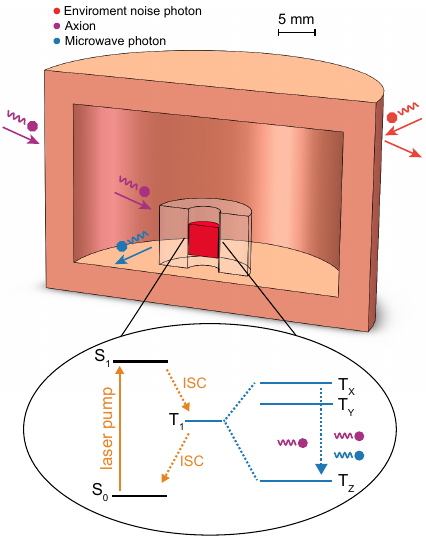}
\caption{\label{FIG.1}
\textbf{Three-dimensional cross-sectional model of the axion detector.} Pentacene-doped p-terphenyl (red) is embedded within a strontium titanate ring (transparent), enclosed in a shell made of oxygen-free copper (yellow), forming a sealed volume. The metallic enclosure provides electromagnetic shielding that blocks external electromagnetic noise photons (red sphere and arrows), while the axions (pink spheres and arrows) can penetrate the copper shell and interact with the photoexcited pentacene molecules. This interaction induces the emission of detectable microwave photons (blue sphere and arrows). Inset: Energy level diagram of the pentacene spin system. After optical excitation, pentacene forms an effective two-level system consisting of the $\rm{T}_{\rm{X}}$ and $\rm{T}_{\rm{Z}}$ sublevels. Interaction with the axion field induces stimulated emission of microwave photons.
}
\end{figure}

The interaction Hamiltonian treats the axion field as an oscillating pseudo-magnetic field oriented along the velocity vector \textbf{v}, with a typical speed $|\textbf{v}| \approx 10^{-3}c$ \cite{PhysRevD.88.035023,PhysRevD.78.063508}. The parameter $\rho_{\rm{DM}}$ denotes the local dark matter energy density, typically taken to be approximately 0.4\,$\rm{GeV/cm}^3$ \cite{Jiang2021}. The vector $\bm{\sigma}_{e}$ represents the electron spin operator. For electron spins, the gyromagnetic ratio $\gamma_{e}$, magnetic moment $\bm{\mu}_{e}$, Landé g-factor $g_{e}$, and Bohr magneton $\mu_{\rm{B}}$ are related by $\gamma_{e}=g_{e}\mu_{\rm{B}}/\hbar$ and $\bm{\mu}_{e}=g_{e}\mu_{\rm{B}}\bm{\sigma}_{e}$. By combining the standard expression $H=-\bm{\mu}_{e}\textbf{B}$ for the interaction energy of a magnetic moment in a magnetic field with the form of the axion-induced pseudo-magnetic field, we derives an approximate relationship between the axion field amplitude and the axion–electron coupling constant:
\begin{align}
     \textbf{B} = 10^{-3}\frac{g_{aee}}{\gamma_{e}\hbar}\sqrt{2\hbar^{3}c^{3}\rho_{\rm{DM}}}sin(2\pi\nu_{a}t)\frac{\textbf{v}}{|\textbf{v}|}\label{eq.3}\,.
\end{align}
After projecting the axion velocity onto the laboratory coordinate system (see Supplemental Materials), we can obtain the maximum measurable value of the pseudo-magnetic field:
\begin{align}
     |\textbf{B}| = 10^{-3}\frac{g_{aee}}{\gamma_{e}\hbar}\sqrt{2\hbar^{3}c^{3}\rho_{\rm{DM}}}sin(2\pi\nu_{a}t)\label{eq.4}\,.
\end{align}

We constructed an axion detector based on maser, as illustrated in Fig.\,\ref{FIG.1}. The  maser is the microwave analog of a laser, which amplifies signals through stimulated emission \cite{PhysRev.99.1264}. It typically consists of a microwave resonator formed by an external metallic shell and an internal dielectric material, as well as a gain medium. The gain medium provides a pair of energy levels with population inversion, and the resonator is tuned to match the transition frequency of the gain medium, enabling stimulated emission and thus producing a strong microwave output. In our implementation, a pentacene-doped \textit{p}-terphenyl crystal \cite{Oxborrow2012} serves as the gain medium, and strontium titanate (STO) \cite{Breeze2015} is employed as the dielectric material. Upon pulsed excitation with 590\,nm yellow laser light, the excited molecules undergo efficient intersystem crossing with a quantum yield of 62.5\% \cite{10.1063/1.1499124} to form a triplet excited state $\rm{T}_{\rm{1}}$. The zero-field splitting between the $\rm{T}_{\rm{X}}$ and $\rm{T}_{\rm{Z}}$ sublevels of $\rm{T}_{\rm{1}}$ is approximately 1.45 GHz, and the initial population ratio between $\rm{T}_{\rm{X}}$ and $\rm{T}_{\rm{Z}}$ is about $0.76$:$0.08$, corresponding to a spin polarization of $\sim 80\%$. By adjusting the quality factor ($Q$) of the resonator, the maser can be operated in an amplifier mode \cite{Breeze2015,zhang2024ultrasensitivesolidstateorganicmolecular}, in which microwave output is generated only in response to an external input signal. The resonator shell is fabricated from oxygen-free copper (99.99\%), whose thickness ($\sim 4$\,mm) is much greater than the microwave skin depth at this frequency ($\sim 1.7$\,$\mu$m), thus providing effective shielding against external electromagnetic interference. In contrast, the axion-induced pseudo-magnetic field is not subject to this shielding \cite{PhysRevD.94.082005} and can penetrate the cavity freely, interacting with the pentacene spins and inducing stimulated emission from $\rm{T}_{\rm{X}}$ to $\rm{T}_{\rm{Z}}$, thereby producing an observable signal.

\section{Experimental results}
\begin{figure}[htbp!]
\centering
\includegraphics[width=8.8cm]{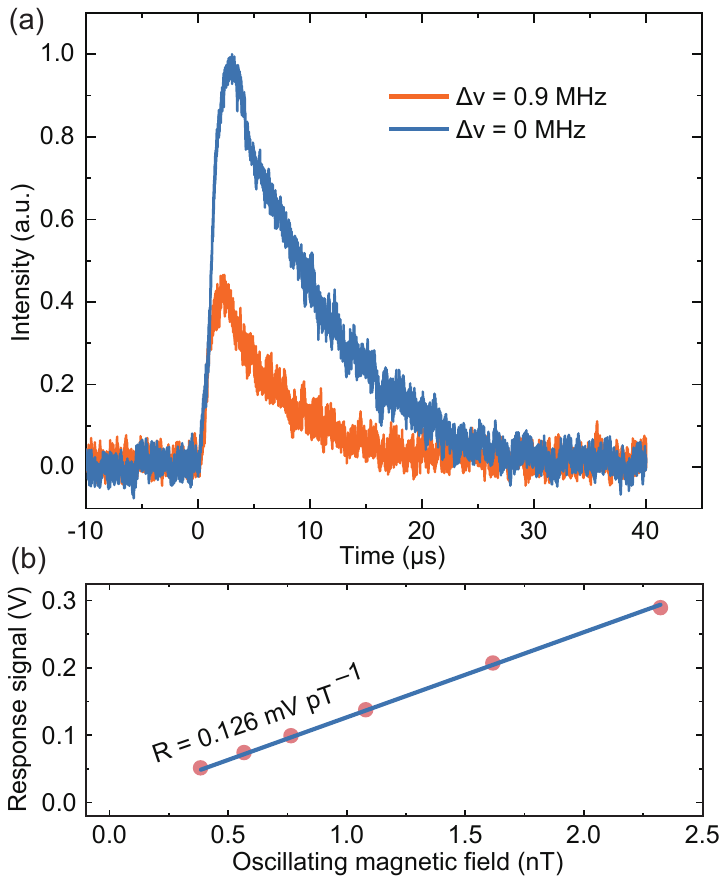}
\caption{\label{FIG.2}
\textbf{Magnetic field response of the pentacene maser.} (a) Single-shot time-domain output signals of the pentacene maser response to microwave magnetic field under resonant ($\Delta\nu = 0$\,MHz, blue) and near-resonant ($\Delta\nu = 0.9$\,MHz, yellow) conditions. (b) At spin resonance, the output signal amplitude exhibits linear dependence on the amplitude of the applied microwave magnetic field. A linear fit yields a slope of $R = 0.126\,\rm{mV/pT}$, which characterizes the system responsivity.}
\end{figure}

\begin{figure}[htbp!]
\centering
\includegraphics[width=8.8cm]{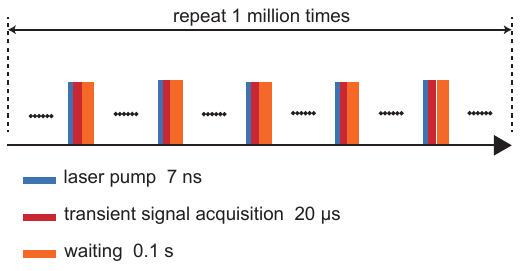}
\caption{\label{FIG.3}
\textbf{Workflow of dark matter detection.} A single search cycle consists of four sequential steps: laser excitation of the sample, time-domain signal acquisition, and a subsequent waiting time because of laser repeat frequency 10 Hz. To ensure consistent of system conditions, the same procedure is repeated 1 million times without altering the experimental setup between cycles.}
\end{figure}

\begin{figure*}[htbp!]
\centering
\includegraphics[width=1\textwidth]{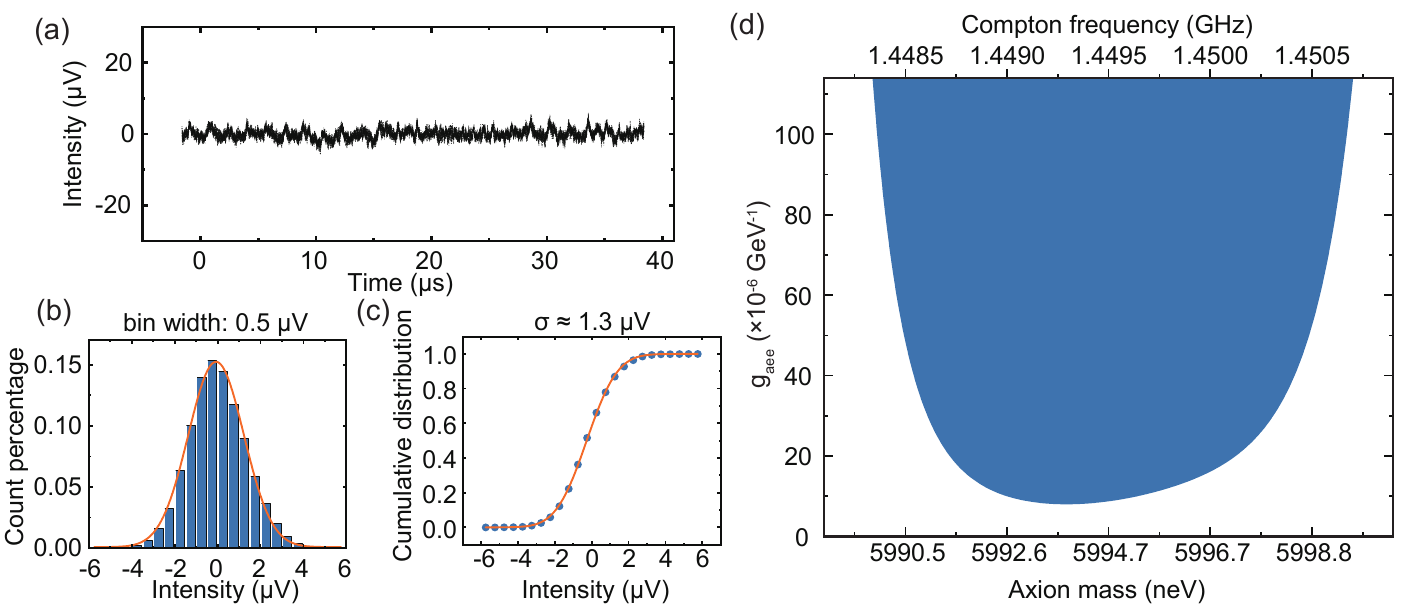}
\caption{\label{FIG.4}
\textbf{Data analysis for axion detection}. (a) Averaged time-domain signal obtained from 1 million measurement cycles. (b) Histogram of the time-domain data with a bin width of 0.5 $\mu$V. A Gaussian fit (yellow line) reveals that the signal follows a standard Gaussian distribution, indicating that the observed signal is white noise. (c) Fitted CDF of the Gaussian noise, yielding a noise standard deviation of $\sigma = 1.31(1)$\,$\mu$V. (d) Experimental constraint on the axion-electron coupling constant at the 95\% confidence level within a mass range centered at $m_a$ = 5993.8\,neV. This constraint is derived using the maser at a resonant frequency of approximately $\nu = 1.4493\,\rm{GHz}$.}
\end{figure*}

To simulate the axion-induced pseudo-magnetic field, we applied a continuous weak high-frequency microwave signal to the cavity, matched to the spin-transition energy. The relationship between microwave power and magnetic field strength was calibrated using standard transient electron paramagnetic resonance (trEPR) techniques \cite{doi:10.1126/sciadv.aaz8244,https://doi.org/10.1002/advs.202401904}. Upon laser excitation, the pentacene molecules are predominantly in the $\rm{T}_{\rm{X}}$ state, so that there is a high population excess over the $\rm{T}_{\rm{Z}}$ state. When a resonant magnetic field is applied, it couples these sublevels, generating a transient pulsed signal. This signal is converted into a DC offset by a logarithmic detector and acquired with an oscilloscope. As shown in Fig.\,\ref{FIG.2}(a), a measurable pulsed signal lasting tens of microseconds is observed immediately after laser excitation at $t$ = 0. Furthermore, when the frequency of the applied oscillating magnetic field matches the spin resonance frequency, the signal intensity reaches a maximum. Figure\,\ref{FIG.2}(b) shows that, with a weak input magnetic field, the output signal intensity exhibits a linear dependence on the magnetic field strength. The slope of this response, i.e., the responsivity, is determined to be $R$ = 0.126\,mV/pT. With the responsivity determined, we quantify the detection sensitivity. The sensitivity is defined as $S=\frac{\sigma_{\rm{noise}}}{R\sqrt{2\Delta f}}$ \cite{doi:10.1126/sciadv.ade1613,PhysRevApplied.10.034044}, 
where $\sigma_{\rm{noise}} \approx$ 3.4\,mV represents the standard deviation of noise in a single measurement, and $\Delta f$ = 500\,MHz is the measurement bandwidth. Accordingly, we obtain a sensitivity of $S= 0.85\,\rm{fT/\sqrt{Hz}}$. The achieved sensitivity is an order of magnitude higher than that of advanced nuclear spin-based axion sensors \cite{Jiang2021,doi:10.1126/sciadv.abi9535,PhysRevLett.129.051801}, yet the maser-based approach provides a conceptually distinct and experimentally accessible route for axion detection. 

Within the widths of the $\rm{T}_{\rm{X}}$ and $\rm{T}_{\rm{Z}}$ sublevels in the triplet excited state ($\sim\,$1.2\,MHz, see Supplemental Materials), we carried out a search for axion signals. During the measurements, we first fixed the resonant cavity frequency and matched it to the spin transition frequency at approximately 1.4493\,GHz. A frequency-stabilization module (see Supplemental Materials) was employed to maintain long-term stability of both the cavity resonance and the spin-environment temperature. As illustrated in Fig.\,\ref{FIG.3}, a total of $N=1$ million measurement cycles were performed. After $N$ repeated measurements, the noise floor can be effectively reduced by a factor of $\sqrt{N}$. In principle, there is no upper limit on the number of detection cycles, although the total measurement time increases proportionally.

After completing 1 million measurements, the acquired time-domain data were averaged, as shown in Fig.\,\ref{FIG.4}(a). To assess the minimum detectable signal strength at a 95\% confidence level, the time-domain data were further processed. Fig.\,\ref{FIG.4}(b) presents the amplitude histogram derived from the time-domain data, with a bin width set to 0.5\,$\mu$V. Analysis of the fitted curve indicates that the histogram follows a standard Gaussian distribution, suggesting that the original time-domain data consist of Gaussian white noise. Subsequently, to obtain an accurate estimate of the standard deviation, the cumulative distribution function (CDF) was plotted in Fig.\,\ref{FIG.4}(c), from which the standard deviation $\sigma =$ 1.31(1)\,$\mu$V. As previously discussed, the axion signal observed on the oscilloscope corresponds to a single-sided DC offset signal; therefore, a single-sided 95\% confidence level test based on the standard Gaussian distribution was applied, yielding the detection threshold 1.645$\sigma$. Here, we should note that the unit of threshold is volt and can be converted into magnetic field according to the responsivity. Further, to obtain the constraint on the coupling parameter $\rm{g}_{aee}$, we can use Eq.\,\ref{eq.4} and the threshold to achieve the corresponding $|g_{\rm{aee}}|\approx 8 \times 10^{-6}\,\rm{GeV}^{-1}$ at central frequency, along with the detection threshold across the full bandwidth, as shown in Fig.\,\ref{FIG.4}(d).

\section{Conclusion}
 In this work, we performed a novel search for axion dark matter based on the pentacene electron-spin maser. Within the zero-field splitting range of the system, we explored the mass scale of approximately 6\,$\mu$eV with a sensitivity to the pseudo-magnetic field of $0.85\,\rm{fT/\sqrt{Hz}}$, thereby complementing the detection limits for axion-electron coupling. This scheme offers advantages including operational simplicity, room-temperature operation without the need for strong magnetic fields, and a compact device structure, demonstrating the feasibility of detecting low-mass axion particles under ambient conditions. Although the current implementation employs a pulsed maser mechanism, future developments toward continuous-wave (CW) operation \cite{PhysRevApplied.14.064017} hold promise for more stable and long-duration dark matter searches. Furthermore, while this scheme is based on the pentacene system, it exhibits excellent scalability and can be extended to other magnetic resonance spin systems with suitable excitation dynamics, such as ruby \cite{2008lnsd.book.....R}, NV centers in diamond \cite{Jin2015,Breeze2018}, and silicon carbide \cite{Kraus2014,PhysRevApplied.9.054006}, thereby addressing detection requirements across different mass ranges or coupling scenarios. However, the maser-based axion detection scheme is intrinsically a highly resonant technique, providing gain only within a bandwidth determined by the cavity mode and the spin-transition linewidth, which limits the accessible region of the axion parameter space. Fortunately, by applying controllable external magnetic fields and exploiting the Zeeman effect to tune the zero-field splitting between spin levels \cite{https://doi.org/10.1002/advs.202401904}, the resonant frequency range of the system can be effectively broadened to cover a wider range of axion masses. By integrating multiple materials with magnetic field tuning in a CW maser platform, this approach shows strong potential for realizing broadband, room-temperature, and cost-effective axion detectors.

\vspace{10pt} %
\section{acknowledgments}
We sincerely thank Min Jiang for the initial inspiration of this work and for the stimulating discussions. We also thank Lei Cong for valuable discussions. This study was supported by NSF of China (Grant No.\,T2522007, No.\,12441502, No.\,12374462, No.\,12321004), the Beijing National Laboratory for Condensed Matter Physics (Grant No. 2023BNLCMPKF007), the National Key R\&D Program of China (Grant No. 2018YFA0306600), by the Cluster of Excellence 'Precision Physics, Fundamental Interactions, and Structure of Matter' (PRISMA++ EXC 2118/2) funded by DFG within the German Excellence Strategy (Project ID 390831469), and by the COST Action within the project COSMIC WISPers (Grant No. CA21106).

\nocite{*}

%

\end{document}